\documentclass{article}

\usepackage{arxiv}

\usepackage[utf8]{inputenc} 
\usepackage[T1]{fontenc}    
\usepackage{hyperref}       
\usepackage{url}            
\usepackage{booktabs}       
\usepackage{amsfonts}       
\usepackage{nicefrac}       
\usepackage{microtype}      
\usepackage{graphicx}
\usepackage{xcolor}
\usepackage{multirow}

\usepackage{doi}

\usepackage{algorithm}
\usepackage{algorithmic}
\usepackage{amsmath}
\usepackage{tabularx}
\usepackage{subcaption}
\usepackage{eqparbox}
\usepackage{comment}

\usepackage{textcomp}

\usepackage{url}
\usepackage{framed}
\usepackage{float} 

\usepackage{changepage}
\usepackage{tcolorbox}
\usepackage{enumitem}
\usepackage{dirtytalk}
\usepackage{balance}

\title{Experiences with Remote Examination Formats in Light of GPT-4
}

\author{
  Felix Dobslaw, Peter Bergh \\
  Dept.\ of Communication, Quality Management and Information Systems \\
  Mid Sweden University \\
  Östersund\\
  \texttt{{felix.dobslaw, peter.bergh}@miun.se} \\
}

\begin{document}
\maketitle

\begin{abstract}
Sudden access to the rapidly improving large language model GPT by open-ai forces educational institutions worldwide to revisit their exam procedures. In the pre-GPT era, we successfully applied oral and open-book home exams for two courses in the third year of our predominantly remote Software Engineering BSc program. We ask in this paper whether our current open-book exams are still viable or whether a move back to a legally compliant but less scalable oral exam is the only workable alternative. We further compare work-effort estimates between oral and open-book exams and report on differences in throughput and grade distribution over eight years to better understand the impact of examination format on the outcome. Examining GPT v4 on the most recent open-book exams showed that our current Artificial Intelligence and Reactive Programming exams are not GPT v4 proof. Three potential weaknesses of GPT are outlined. We also found that grade distributions have largely been unaffected by the examination format, opening up for a move to oral examinations only if needed. Throughput was higher for open-book exam course instances (73\% vs 64\%), while fail rates were too (12\% vs 7\%), with teacher workload increasing even for smaller classes. We also report on our experience regarding effort. Oral examinations are efficient for smaller groups but come with caveats regarding intensity and stress.

\end{abstract}


\section{Introduction}

Reliable knowledge assessment is essential for educational institutions to be regarded as quality education providers. The Swedish Higher Education Authority (Universitetskanslersämbetet, UKÄ) defines rules for legally compliant examination in Sweden\footnote{\url{https://www.uka.se/swedish-higher-education-authority}}, but each university has the liberty to decide suitable examination formats. A proper design must guarantee authentication and ensure fair as well as reliable assessment.

Large Language Models (LLM) recent and primarily free of charge availability to the greater public, with the Generative Pre-trained Transformer (GPT) web interface ChatGPT at the forefront, now call the reliability of assessment formats in question. The quality of these general models, both in terms of language and seeming comprehension, raises the concern of whether students could proxy questions and copy the answers to submit them in their name. These deep learning-based AI engines are seemingly creative and can be communicated with in conversations. However, they retain a memory of previous comments, making it hard to assess the future trajectory of this technology, considering that its broad data collection has just begun. LLMs are projected to substantially impact the labour market as a whole, with professions in the software industry being high up on the list \cite{eloundou2023gpts}, which affects us professional software educators.

Therefore, we investigate how GPT version 4 (GPT-4) performs to understand vulnerability in our third-year BSc software engineering exams. We do this in light of our vast experience with oral exams, which are not as easily breached. With empirical data on grades and work effort in oral and open-book home exams, we more broadly analyze the various implications assuming expanding challenges associated with remote exams in the future. This is particularly important for our education, as it is predominantly studied by students that are 100\% remote. Furthermore, entirely supervised written exams have logistically been challenging to arrange and reduce the program's attractiveness due to reduced student convenience.

The paper is structured as follows. Section \ref{sec:background} describes the relevant formats for examination in our studies and surveys the use of LLMs in education. Section \ref{sec:method} presents the research questions, gives an overview of our educational context, and details the data collection. In Section \ref{sec:discussion}, we present our results in response to the research questions and bring up threats to the validity of this study. Section \ref{sec:conclusion} concludes the paper with an outlook into future work.
\section{Background}
\label{sec:background}

While assessment in higher education today is predominantly conducted through written exam formats, oral exams (or examination interviews) have a long history dating back to the origins of educational institutions in the West \cite{stray2001shift}. Pros and Cons exist for both formats~\cite{gharibyan2005assessing}. Some studies suggest oral exams lead to better performance, and one possible explanation is the direct conversational nature which may make students prepare harder to avoid embarrassing themselves in front of the professor \cite{huxham2012oral}. One advantage of written exams is that they are more easily made to scale in the number of students, and parts of the evaluation can be automated or augmented~\cite{paiva2022automated}. Another aspect to consider is that, while the demand for communication skills is rising on the job market \cite{scaffidi2018employers}, the written examination does not seem to be a suitable format for assessing it, as a large proportion of communication is happening non-verbally, possibly as much as 93\% \cite{lapakko2007communication}, and the theoretical knowledge may not directly translate to the practical competency to be evaluated. Oral exams spread throughout a program can offer good training opportunities instead of written ones.

During the Covid19 pandemic, when people could not meet, a move over to remote examination was forced. Written home exams, often under proctoring, were prominent choices~\cite{hatzipanagos2020towards}, but even oral exams were heavily applied and thoroughly evaluated (see, e.g., \cite{graf2021online}). Having students do an exam in front of the computer from home substantially simplifies cheating compared to on-site written and supervised exams. However, these so-called take-home exams come with a risk of access to external support. With remote supervision and otherwise stringent time constraints, these risks can be mitigated to some extent. \textit{Proctoring} somewhat mitigates the risks for cheating, with students needing to remain logged into a system for the exam and to share a camera view for the teacher \cite{alin2022addressing}. Just like their on-site counterpart, take-home exams can be open-book or closed-book. The open-book approach usually allows students to use all literature, including the Internet, to get access to information, as long as no consolidation (previously to humans only, now even to LLMs) is taking place. The closed book format entirely disallows the use of any external resources - a design much harder to enforce in a remote setting. The challenge for open-book exams, then, is to ask questions that cannot quickly be answered directly from the literature but require a transfer of knowledge to the task at hand, which puts greater demands on the creativity of the teacher, further requiring questions to change continuously as it is easier for students to record questions for later exams. Alin et al. describe this challenge under the term \textit{Assessment design}, offering guidance in \cite{alin2022addressing}. The other two strategies they synthesize for dealing with cheating in take-home exams are monitoring and limiting resources.

However, there is no bulletproof solution. While combining different measures was advised but cumbersome already in a pre-GPT era, the availability of tailored machine-based LLMs potentially made it necessary. The alternative, with which Alin et al. close, captures the current situation educational institutions are in: \say{\textit{...if assessments are becoming prone to cheating due to technological advances, and if mitigating cheating is ethically or practically impossible, it may be time to rethink how to best assess students in the future...}}\cite{alin2022addressing}. In that vein - are oral exams a (the only remaining) workable alternative for remote education programs?

Large Language models (LLM) apply Natural Language Processing (NLP), i.e. they respond in natural language to natural language instructions\slash questions - usually through a prompt or API. While early LLMs were problem-specific, such as Github Copilot or Codex \cite{chen2021evaluating} for programming tasks, did the enrolment of the general-purpose GPT model receive massive attention because of its (multi-)language skillfulness but also obvious flaws on a general\slash broad knowledge spectrum. What further sets ChatGPT\slash GPT apart from other models is its capability to interact in more extended conversations regarding previous comments, as opposed to the one-shot use of, e.g. GitHub Copilot. This feature has found its way into search engines and may redefine search altogether. Another feature included in v4 of GPT is that it can use images as input\footnote{Image use was though not available to us when we used the initial versions for this study.}.

With their great potential as work assistants, LLMs are investigated for educational uses and harm to legally compliant examination as we know it \cite{kasneci2023chatgpt}. For programming, Github Copilot, for instance, seems to produce comprehensible enough results to create immediate value for newcomers, while the code produced still differs from what humans usually write \cite{puryear2022github}. However, because public availability is limited (recently enrolled and added cost of use), little research has been spent on how to use LLM in education.

Suppose the solution is between moving to oral examinations or necessary format adjustments in a remote exam. In that case, there is a need for more empirical evidence through case studies that help understand practical and safe examination implications.
\section{Method}
\label{sec:method}

While a large body of literature on experiences with LLMs can be expected, so far, there is little practical guidance on using them in this new era of ChatGPT and LLMs availability. At the same time, there is a lack of studies comparing oral and open-book examinations quantitatively regarding grade distributions and throughput as decision support. We, therefore, ask the following research questions to address that gap:
\begin{enumerate}
    \item How does the examination format (open book home exam vs oral) impact grade distribution and throughput?
    \item How do the examination formats compare in terms of examiner workload?
    \item How does GPT-4 perform in answering third-year programming-heavy computer science exam questions?
    \item Can question types or formats be identified for which GPT-4 struggles?
\end{enumerate}

We address RQ1 by direct statistical comparison over the grade distributions between oral and open-book home exams. Similarly, throughput, or student passing ratio, is compared as an aggregated mean among all course instances. Throughput is an essential measure in Sweden as it impacts the state reimbursement to the university (each passing student produces income). RQ2 is addressed by rough experience assessments over the open questions of work effort required for preparation, conduct, and grading (incl. reporting) to get approximate numbers for four scenarios of 1, 30, 60 and 100 students. The assumption is that previous exam material exists, i.e. we focus on maintenance, and this is not the very first exam for a new course. To answer RQ3, we asked ChatGPT in a one-shot style about the answers to all exam questions and graded the result manually, just as we would have with a student. Finally, RQ4 is answered based on the experiences from RQ3.

\subsection{The Studies Context}
The study is based on two five-week full-time equivalent courses (7.5 ECTS credits) taught in the third year of a BSc degree in Software Engineering at a Swedish University. The program is hybrid, with on-site but predominantly remote students. The course subjects are Reactive Programming (RP) and Artificial Intelligence (AI), both having a lab part with programming deliverable(s), as well as a final exam to assess the learning outcomes. Grades for the labs are \textit{pass} and \textit{fail}, and for the exam, according to Bologna, A-F (A-E for passing, and F for fail). The exam grade defines the course grade. All on-site and remote students are examined under the same conditions, defaulting to a remote setting.

The RP course started in 2019, and we have grading data from four course instances - two examined through oral exams (2019-2020) and two through open-book home exams (2021-2022). For the AI course, we have data from eight course instances - six examined through oral exams (2015-2020) and again two through open-book home exams (2021-2022). Our total data, therefore, persists in eight course instances for oral and four for written exams. The format change was due to a change in responsible teacher and personal preferences only but in close teacher collaboration\footnote{Both teachers and the course examiner are co-authors of this paper.}. All students are eligible for any grade based on individual performance - no relative grade distribution is applied.

\subsection{Examination Conduct}
\subsubsection{Oral exam}
Interviews take place via a conferencing system, requiring student audio and video. All examinations are recorded as official documentation - this can be compared to the pages kept as records in a traditional written exam. 30-minute slots are announced in time before the examination day(s) on a first-come, first-serve sign-up basis. For preparation, guidelines are shared with the students concerning the procedure. An interview lasts 20 minutes and ends with a 10-minute reflection round. Throughout the interview, the examiner asks questions to cover the relevant areas and, at times, in support of, e.g. figures or code, takes notes, and the grade is usually reported the same day. The examiner has several alternating questions for each of the usually 6-8 areas to be covered. Questions vary in the order and way they are asked and are explored in response to the received answers.

\subsubsection{Open-book home Exam}
The sign-up procedure is similar to regular on-site exams. The RP exam lasts 60 minutes and consists of 13 questions, with a maximum of 38 points that can be obtained (19 points are required to pass). For AI, the exam lasts 150 minutes and consists of 26 questions, and a total of 68 points can be obtained (29 points are required to pass). Students access the exams through the learning platform from their home computers. They must throughout be available on a video call where questions could be asked to\slash by the teacher, but disallowing peer- communication.

\subsection{Data Collection}
\label{datacollection}
We collected historical student records and LLM answers for two exams.

The public student records were collected through the look University database for all twelve course instances. The records were downloadable in CSV format and contained information about registration, completion, repeated attempts, and grades and dates. All grading information was anonymized - personal identifiers were substituted with unique numbers per course, and the date information was transformed into the respective year.

We used ChatGPT v4 (paid tier), with a simple (one-shot) approach to collect the exam questions from two open-book exams for the current instance of both courses. We considered the single direct answer as the final answer for the grading done by the responsible teacher, just like for any other student. All questions but one\footnote{From the AI exam, noted as OMITTED in the reproduction package.} could be processed, while those based on illustrations required a written description instead. Our natural language descriptions of the illustrations are in the shared reproduction package.
\section{Results and Discussion}
\label{sec:discussion}

We here answer the research questions one at a time and finish the section with key takeaways in \ref{takeaways} and Threats to validity in \ref{threats}. All anonymized grading information, student answers, exam questions, student answers, and code for reproducing the statistics and plots are available online\footnote{\url{https://1drv.ms/u/s!AuZALvfcrtYImfpIrRyJiFN38darhA?e=ZDxucN}. 
}.

\subsection{RQ1: Quantitative Differences}

\begin{figure}
    \centering
    \includegraphics[width=\columnwidth]{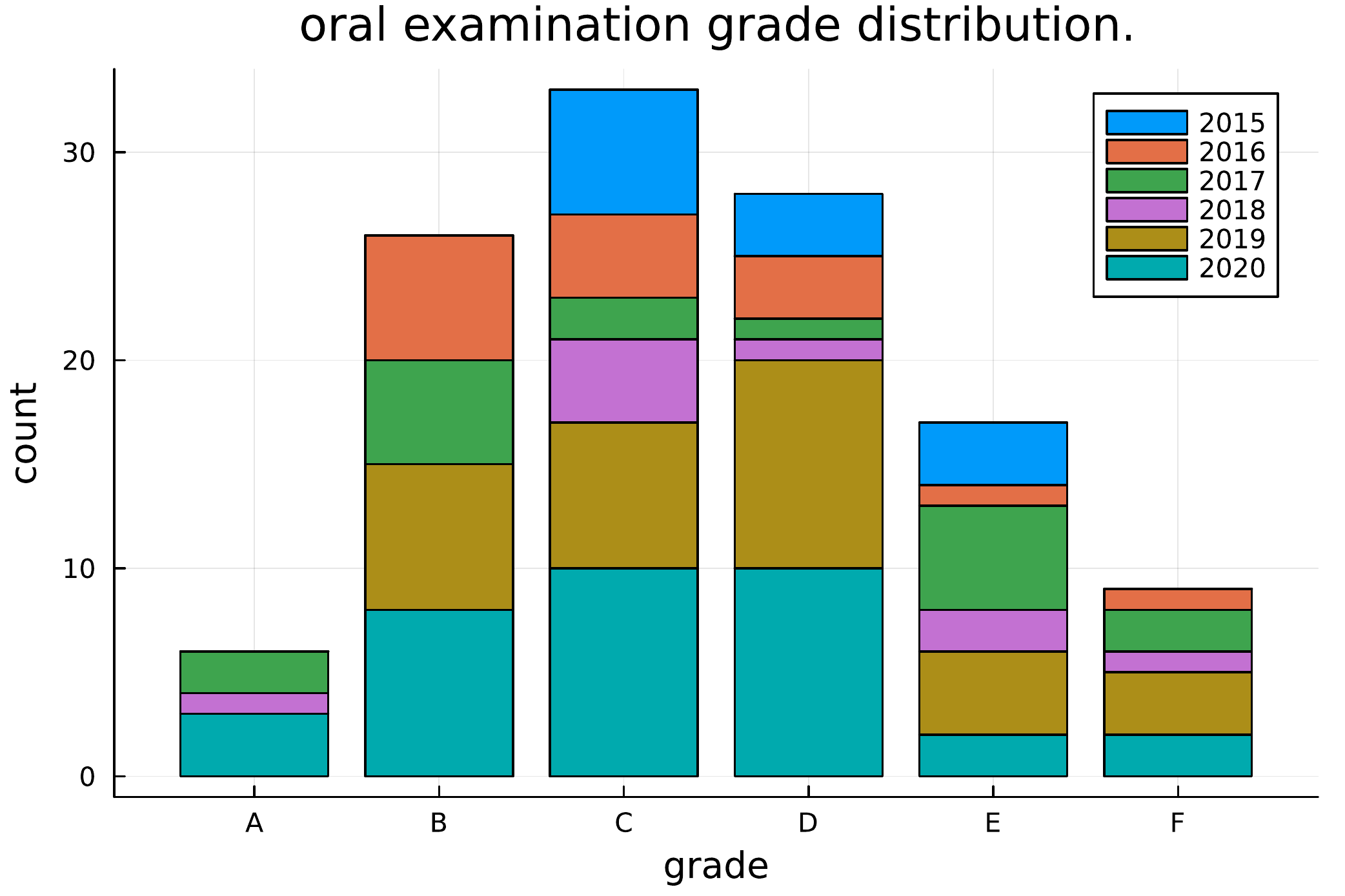}
    \caption{The grade distribution over all oral exams (eight course instances, 2015-2020).}
    \label{fig:oral}
\end{figure}
\begin{figure}
    \centering
    \includegraphics[width=\columnwidth]{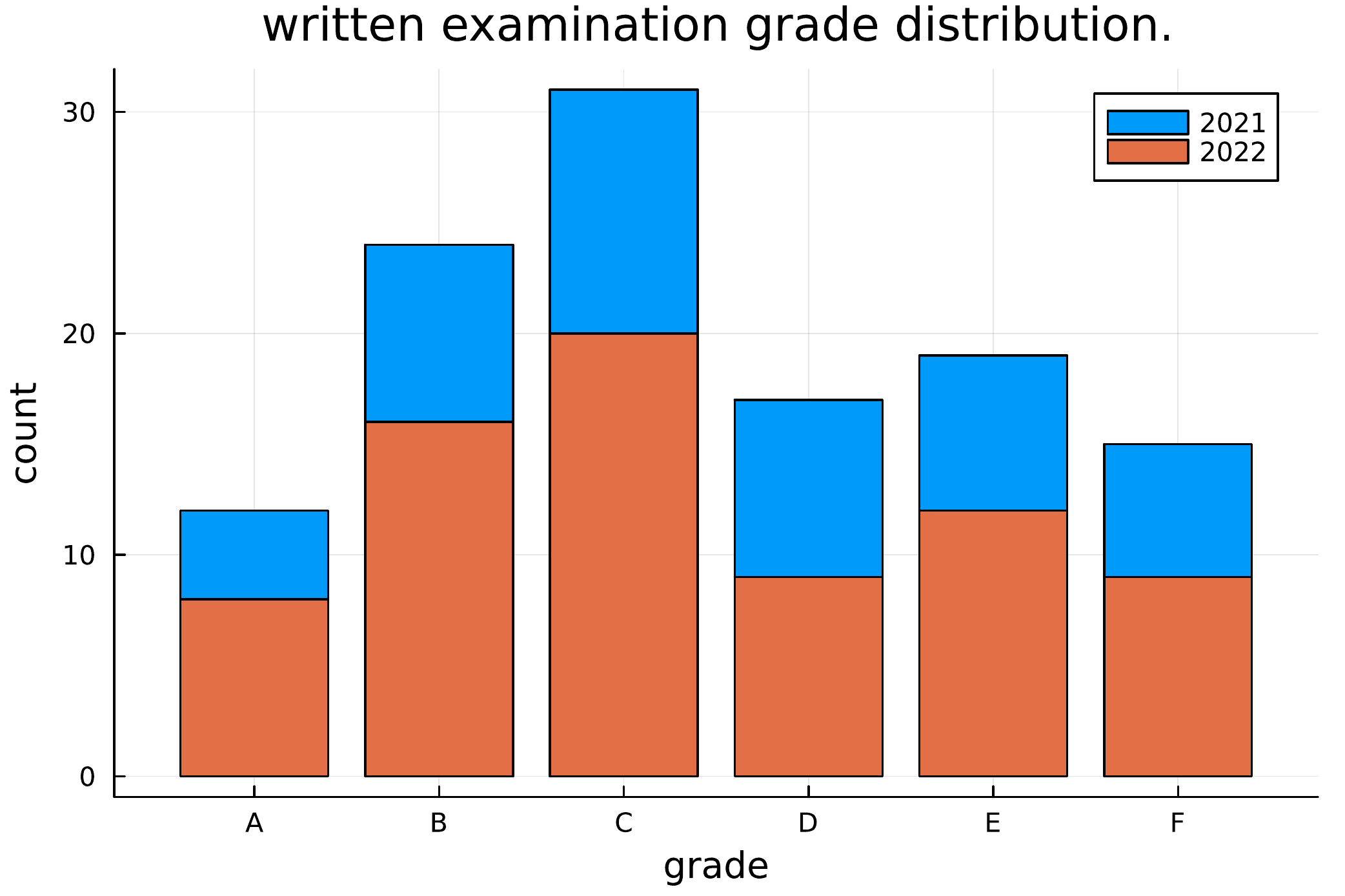}
    \caption{The grade distribution over all open-book exams throughout the years (four course instances, 2021-2022).}
    \label{fig:written}
\end{figure}

Figure \ref{fig:oral} shows the grade distribution for all oral examinations, and Figure \ref{fig:written} the grade distribution for all written examinations. All oral examination attempts were 119 and written ones 118. In later years, we experienced a rise in student numbers.

The oral exam distribution looks approximately Gaussian, the written exam one slightly less so. We did a Mann-Whitney U test, a non-parametric statistical test, to see whether the distributions follow the same central tendency (null hypothesis). The data is ordinal, which is why this ranked-based test is preferred to, e.g. the Student-t test. The test reports with a U value of 18.5 that the null hypothesis could not be rejected, i.e. the distributions still may follow the same central tendency at a 95\% confidence level and a two-sided p-value of 1.




For the oral examinations, we had a total of 165 students, of which 107 or 64\% completed the entire course, including labs. For the written exam, there were 131 students in total and 96 passing, thus 73\% throughput. Thus, throughput may be affected by the format, in our case, with higher rates for written exams. However, the overall fail rates for the exams were with 9\slash 119 (7\%) lower for oral exams than for open-book exams (14\slash 118 or 12\%), suggesting that the exam format in itself may not be the causing factor.

\subsection{RQ2: Workload Differences}
Table \ref{tab:diff} compares the three scenarios regarding minutes spent and based on personal experience. We focus on reporting things that do not have to be done commonly otherwise (e.g. sending announcements for the sign-up or reminders have to be done in both situations). We did not either include time for students challenging the decision, as we assume here that this is equally distributed in time and effort. The experience on oral examinations is that only two students challenged the decision throughout the six years. 

\begin{table}[]
\caption{The assessed effort for grading oral vs open-book home exams (obhe).}
\small
\centering
\begin{tabular}{lrcccc}
\label{tab:diff}
\begin{tabular}[c]{@{}l@{}}number\\students\end{tabular}&\begin{tabular}[c]{@{}l@{}}exam\\format\end{tabular}&\begin{tabular}[c]{@{}c@{}}preparation\\(mins)\end{tabular}&\begin{tabular}[c]{@{}c@{}}conduct\\(mins)\end{tabular}&\begin{tabular}[c]{@{}c@{}}grading\\(mins)\end{tabular}&\begin{tabular}[c]{@{}c@{}}total\\(hours)\end{tabular}\\
\hline
1&oral&\textbf{180}&\textbf{30}&\textbf{2}&\textbf{3.5}\\
&obhe&480&60-150&20&9.3 - 10.8\\
\hline
30&oral&\textbf{180}&900&\textbf{22}&\textbf{18.4}\\
&obhe&480&\textbf{60-150}&600&19 - 20.5\\
\hline
60&oral&\textbf{180}&1800&\textbf{44}&33.7\\
&obhe&480&\textbf{60-150}&1200& \textbf{29 - 30.5}\\
\hline
100&oral&\textbf{180}&3000&\textbf{74}&54.3\\
&obhe&480&\textbf{60-150}&2000&\textbf{42.3 - 43.8}
\end{tabular}
\end{table}

\subsubsection{Oral exam} Preparation is assessed as three hours and includes the handling of the booking (making times available through a tool) and extends to the updating of last year's question slides in accordance with course updates overall. The conduct is 30 minutes per student which includes the 20 minutes interview as mentioned above. Grading is done after an examination day with seven hours of examination (14 students), resulting in a total of 10 minutes reporting.

\subsubsection{Open-book home exam} To distinguish the exam content from year to year, a working day is spent on preparations. During the exam, the procedure is monitored via video call. An average of 20 minutes is spent checking a student's submitted exam, regardless of course. Both RP and AI exams take around 20 minutes per student to grade. Although the AI exam has twice as many questions as the RP exam, only a quarter require manual grading. In contrast, all questions in the RP exam require manual grading. However, roughly half of the answers in RP are of the type with only one correct solution, which reduces the grading time for those questions. Despite the differences in the number and type of questions between the two exams, the time required for grading each student is thus similar. Reporting grades can be disregarded as a time factor, as this one-time occurrence takes about 10 minutes.

\begin{figure}
    \centering
    \includegraphics[width=.95\columnwidth]{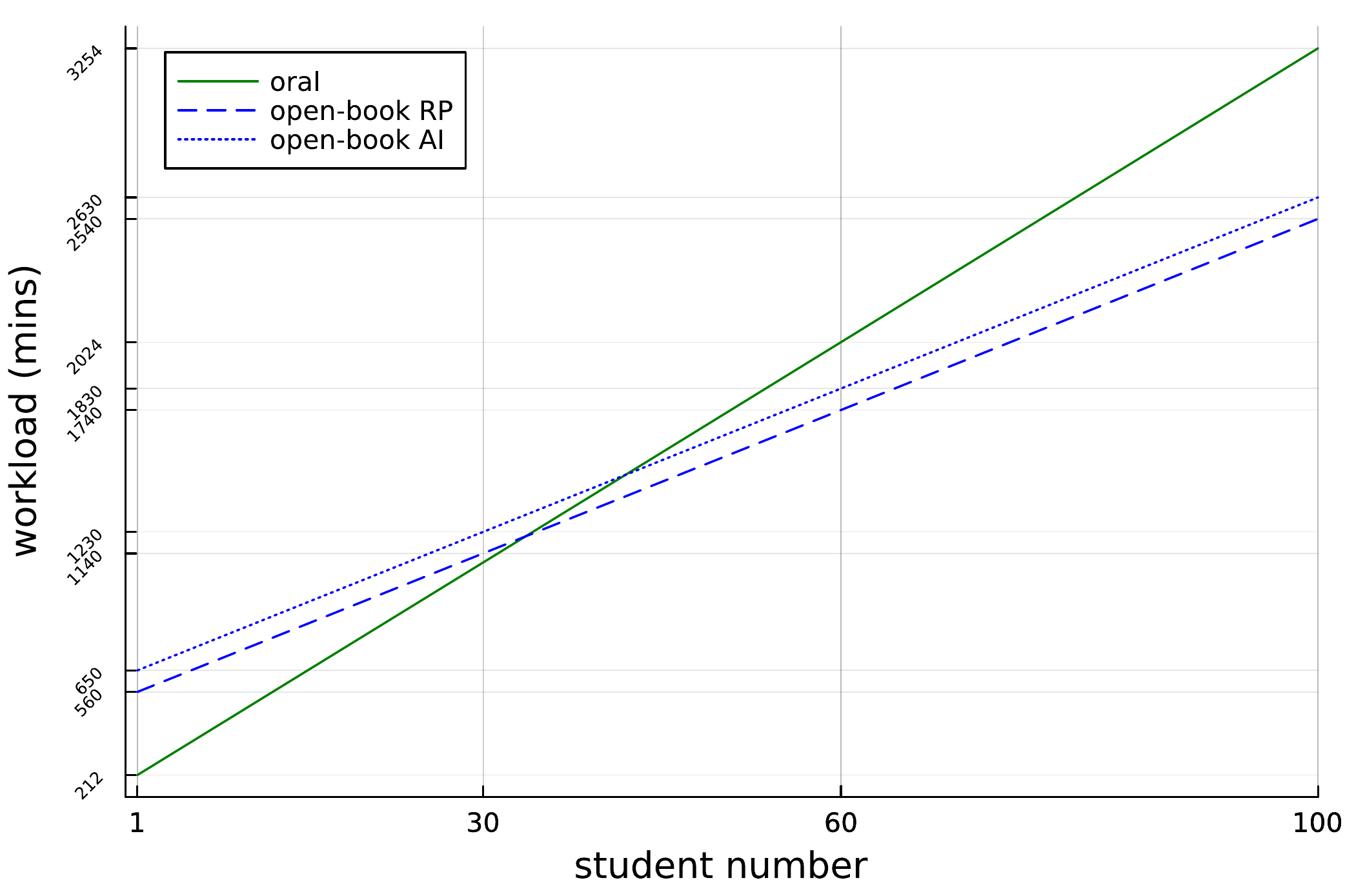}
    \caption{Examiner workload in the number of students and examination format.}
    \label{fig:workload}
\end{figure}

Figure \ref{fig:workload} shows the workload estimation for the respective exam formats in the number of students. For few students, below 30, oral exams are more efficient in our experience. For 40 or more students, the open-book alternative starts to be comparatively more time-efficient.

\subsection{RQ3: Student ChatGPT}
ChatGPT obtained a score of 35\slash 38 points for the RP exam and a score of 40\slash 68 points in AI. This leads to two passing grades, an A for RP (highest grade) and a D for AI. For RP, ChatGPT only fails to answer one question due to what seems to be a misunderstanding in the term \textit{Operator}. Also, one question based on an illustration [question AI-20] could not be described in written form because it contained a state-space landscape which was too complex to describe in written form. ChatGPT demonstrated varying performance in comprehending graphical content described as text, with excellency [e.g. RP-9] as well as struggle [e.g. AI-25].

\subsection{RQ4: ChatGPT v4 question types}
\label{RQ4}
We here offer insight into the types and formats of questions that GPT-4 may struggle with. However, it is worth noting that these observations are primarily based on the AI exam's results. GPT-4 could answer all questions on the RP exam without any issues. Nevertheless, the AI exam provides an opportunity to identify specific question types and formats that may present challenges for GPT-4.

Based on the context and examples provided, GPT-4 appears to struggle with several question types and formats. For example, it may have difficulty with multiple-choice and true/false questions requiring strict adherence to the literature, possibly due to insufficient understanding of the subject matter, misinterpretation of the questions, or the influence of diverse perspectives in its training data.

Regarding multiple-choice questions, GPT-4 often tended to select only some of the correct options or choose too many incorrect ones [AI-11 and AI-2, respectively]. Possible reasons for this struggle could be GPT-4's limited understanding of the nuanced language and phrasing used in the questions, which could have caused confusion or difficulty in identifying the correct answer options.

Despite being a simple multiple-choice question with only one correct answer, GPT-4 answered a question incorrectly [AI-19]. It is possible that the question's context presented a challenge for GPT-4 to comprehend, which may have contributed to the incorrect response. Interestingly, this question is often answered correctly by students.

GPT-4 performed well on essay questions that required reasoning [AI-18] but struggled with questions that required method implementation [AI-14, AI-24, AI-25]. Additionally, GPT-4 had difficulty with graph traversal and search algorithm execution based on textual descriptions of graphs [AI-4, AI-5, AI-6, AI-7]. Mistakes in these questions suggest that GPT-4 may have difficulty in graph visualization and correct search.

In the context of a game tree where nodes were designated as minimizing or maximizing, GPT-4's reasoning process was accurate [AI-12, AI-13]. However, there was a misunderstanding regarding node classification (minimizing vs maximizing), possibly due to deficits in interpreting the game tree description.

\subsection{Takeaways}\label{takeaways}
Regarding the creation of take-home exams in the era of GPT, we have three main observations based on our findings. First, GPT-4 excels in answering open-ended questions where it can reason, describe, and explain, e.g. concepts, algorithms, or weigh different options. At the same time, it can struggle with calculations or solving actual problems, even when perfectly describing a mechanism such as arc-consistency, as mentioned in Section \ref{RQ4}. After all, this should not be surprising as GPT is \textit{only} a language model. We would therefore try to reduce the emphasis on the first category (reasoning, arguing) and focus more on the latter (calculating, solving). Second, describing problems based on illustrations as text can be challenging, especially considering that students may have limited time. That, in particular in the computer sciences, descriptions of graphs, mazes, etc., require some experience and mathematical abilities. We would therefore try to include concrete puzzles or problems (connecting to the first point) that are preferably hard to describe in text. However, since we could not yet test the image recognition capabilities, we cannot speak for the possibilities that might lure just around the corner. Third, multiple-choice questions seemed more challenging for GPT-4 than open-ended ones. This limiting of the alternatives has been shown to lead to surprisingly wrong answers (such as question AI-2 on heuristics improving on solutions), particularly true multiple-choice questions, i.e. where more than one alternatives are correct (and possibly as crucial that multiple ones are wrong). However, this may be a technicality that could be resolved in future versions. Considering all that, we hesitate to recommend offering take-home exams such as the ones investigated in this study, even considering all three points raised.

Making our evaluation and data available makes us vulnerable, as students now know they can cheat their way to success. We, therefore, have a responsibility to make substantial changes before the next course iterations. In our opinion, the alternative, i.e. to pretend this was not a problem and go on as before, is a worse one. We prefer to participate in the overall debate\slash development and update our understanding of the students' and universities' interests to the best of our ability.

An aspect we did not discuss in detail is the re-examinations, with usually much smaller groups. While similar efforts can be expected regarding preparation and conduct for a written exam, for oral exams, no extra preparation may be needed. There, the material is reused in different ways during the interview, i.e. the 180 minutes of preparation might reduce to 30 minutes for the booking of a hand full of slots. For takeaway exams, either way, a new exam must be created for all students.

Finally, we also want to highlight that oral examination is not a format that can be taken from the shelf and applied. They put a high demand on integrity and the communication skills of the examiner, who should be well aware of persuasion techniques and keep a positive attitude towards all students in vulnerable and stressful situations. Special training might be required or advised; not every examiner may feel comfortable with the format. There is no anonymity in oral exams, whereas written ones are traditionally anonymized to reduce interpersonal bias. However, in how we conduct open-book exams, anonymization is not enforced either.

\subsection{Threats to Validity}
\label{threats}
Several threats to the validity of this study exist, which must be discussed. First, the received results from ChatGPT cannot be reproduced because the service is a black box that can change in the background at any time. Second, we could not compare to older versions (e.g. 3.5), as only the latest version is available. This also implies that a struggle to answer questions right may not remain, and ChatGPT may respond entirely differently if attempted later. Third, we did not verify the variance in answers, i.e. asking the questions multiple times and evaluating each solution.

Since oral and open-book exams were conducted by different teachers (all oral by one E1, all written by another E2), there is a risk of individual bias and differences in assessment and preferences. Also, the examiners had varying professional experience (E1, ca. 15 years vs E2, 1-2 years). However, all questions to the initial written RP exam were created by E1, and E2 completed the AI exam based on and inspired by oral exam recordings and material from the former E1. E2 is a former student of E1. Additionally, time estimations for RQ2, although based on two very different courses, do not generalize as they are based on the crude assessment for a limited number of courses.

It should also be noted that the text descriptions of illustrations that ChatGPT could not parse contain an individual bias, and better descriptions may exist. We tried to mitigate this threat by adding all texts to the reproduction package to allow everyone to judge by themselves.
\section{Conclusions}
\label{sec:conclusion}

Given the sudden broad public access to LLMs, we investigated the impact of going \textit{back} to oral exams regarding time exposure and grade distribution. We also tested the urgency of change by examining ChatGPT as a student in two current open-book exams from our BSc program.

Our analysis revealed that the exam format does not seem to have a (statistically) significant impact on the grade distribution. Course throughput for open-book exams was higher (73\% over 64\%) while fail rates were higher too (12\% vs 7\%) in our data from 2015 - 2022. Fewer students attempted oral exams. We further found that ChatGPT passes both exams, one with a top grade. Regarding teacher workload, it should be said that oral examinations are demanding for the examiner concerning integrity and lead to intense long days in dialogue which may affect the impartiality.

As a consequence for us, a change in format\slash implementation of both exams is required. Even if explicitly disallowing ChatGPT\slash GPT, we have no good way to ensure that practically given the current tools and methods \cite{alin2022addressing}. One possible weakness of ChatGPT is the relatively high rate of errors in arithmetic and deduction, which can sometimes be accompanied by excellent explanations (that then contradict the calculations).

Considering the pace of improvement, our experience with ChatGPT, and the empirical evidence, it is hard to see how open-book exams in their current format can be applied even in the near future. Continuous examination with the conversational examination or written on-site exams in physically proctored settings may be required to live up to higher education authority standards around the globe.

In future work, we want to investigate the consequences of GPT's image recognition capabilities once available. Further, we need to understand the impact of one main feature GPT-4 comes with but which we have not exploited in this study - conversation mode. In collaboration with GPT, can we develop and refine answers instead of the one-shot question\slash answer approach? And how about assessment augmentation in support of GPT?


\bibliographystyle{IEEEtran}
\bibliography{bibliography}

\end{document}